\documentclass[11pt,a4paper]{article}
\usepackage[margin=29mm]{geometry}
\usepackage[T1]{fontenc}
\usepackage[utf8]{inputenc}
\usepackage{lmodern,microtype}
\usepackage{amsmath,amssymb,amsthm,mathtools}
\usepackage{booktabs,enumitem}
\usepackage[authoryear,round]{natbib}
\usepackage[colorlinks=true,linkcolor=black,citecolor=black,urlcolor=blue!55!black]{hyperref}
\usepackage{xcolor}
\usepackage{setspace}
\numberwithin{equation}{section}
\newtheorem{theorem}{Theorem}
\newtheorem{proposition}{Proposition}
\newtheorem{lemma}{Lemma}
\newtheorem{corollary}{Corollary}
\theoremstyle{remark}

\newcommand{\qb}{\bar q}
\newcommand{\Xb}{\bar X}
\newcommand{\lb}{\bar\lambda}

\newif\ifanonymous
\ifdefined\anonymized\anonymoustrue\else\anonymousfalse\fi
\hypersetup{pdftitle={Fiscal Commitment and the Additionality of Philanthropy: Heterogeneous Donor Purposes in a Shared Public Budget}}
\ifanonymous
\hypersetup{pdfauthor={}}
\else
\hypersetup{pdfauthor={Christian Jaag}}
\fi
\title{Fiscal Commitment and the Additionality of Philanthropy:\\Heterogeneous Donor Purposes in a Shared Public Budget}
\ifanonymous
\author{}
\else
\author{Christian Jaag\thanks{Swiss Economics, Ottikerstrasse 7, 8006 Zurich, Switzerland, and Somaha Foundation, Weinbergstrasse 102, 8006 Zurich, Switzerland. Correspondence: \texttt{christian.jaag@somaha-stiftung.ch}.}}
\fi
\date{27 October 2026}
\begin{document}
\maketitle
\begin{abstract}
Philanthropic donors often value different public services, while a budget authority allocates public spending across a shared portfolio. Policy frameworks frequently seek to ensure that private finance supplements rather than displaces public funding. We distinguish baseline additionality, which preserves pre-gift public funding, from portfolio preservation, which preserves a service's provision after all gifts are pooled under discretionary budgeting. A larger gift to one service raises the discretionary provision of every service and reduces the protection slack available to the others. This cross-donor fiscal externality yields an exact feasibility result: if a service's maximum permissible floor lies below its discretionary post-gift provision, every effective protection reform reduces that service relative to the portfolio reference, including full protection of its own appropriation. Conversely, a reform that expands at least one donor-backed service while preserving all others exists exactly when every maximum floor reaches the reference. The paper characterizes the complete preservation set and its welfare-maximizing element. Under logarithmic preferences, donation shares determine protectability, while heterogeneous donor valuations determine welfare incidence. Baseline additionality may therefore hold while a service loses fiscal benefits generated by other gifts.
\end{abstract}

\section{Introduction}

Public authorities increasingly interact with philanthropists who direct resources toward particular public services. Donors often value these services differently from the budget authority and from one another, so private finance introduces heterogeneous purposes into a common fiscal system. Policy practice frequently seeks to preserve the additional character of such finance. The United Kingdom's 2026 place-based philanthropy strategy, for example, states that philanthropy should complement public money rather than substitute for state funding \citep{DCMS2026}. Similar principles appear in maintenance-of-effort and supplement-not-supplant requirements for public grants; current U.S. law, for example, requires specified federal education funds to supplement rather than supplant non-federal resources \citep{GAO1980,USCode7120}. These rules share a basic objective: external finance should expand supported activity instead of releasing an equal amount of public funding. The terminology in this paper is model-specific. Policy documents also use ``additionality'' for other principles governing the relationship between government and philanthropy, including donor independence \citep{DCMS2026}.

A shared budget makes this objective more subtle. A gift to one service expands the resources available to the budget authority, which may use the resulting fiscal room across several services. Another donor-backed service may therefore benefit from that gift even when its own public appropriation remains unchanged. A commitment to preserve the second service's pre-gift appropriation protects one reference level, while the service's provision after all gifts have entered the budget provides another.

We distinguish these two references throughout the paper. \emph{Baseline additionality} means that the public appropriation in a donor-backed service is preserved when its own gift arrives. \emph{Portfolio preservation} is stronger: the service retains the provision it would receive after all gifts are pooled and the authority allocates the enlarged budget freely. A full protection rate guarantees baseline additionality in the model. Portfolio preservation may nevertheless fail because the service may lose fiscal benefits created by gifts to other purposes. We use portfolio preservation as a diagnostic policy benchmark rather than as a distributive entitlement. It asks whether an additionality rule is able to preserve a service level already generated by the full portfolio of philanthropic finance; the welfare analysis later treats preservation as an explicit institutional constraint.

The analysis first identifies the fiscal mechanism behind this gap. At the unrestricted post-gift allocation, a marginal increase in any one gift raises provision in every public category. The service receiving the larger gift also raises its own maximum contractual floor one-for-one, whereas the contractual ceilings of the other services remain unchanged. Consequently, a larger own gift expands a service's protection slack, while larger gifts to other purposes reduce it. This cross-donor fiscal externality links heterogeneous philanthropic purposes even before any protection rule binds.

The main theorem then gives an exact feasibility condition. If the maximum permissible floor of any service lies below its discretionary post-gift provision, every effective reform in the permitted class reduces that service relative to the portfolio reference. Differentiated protection rates do not remove the loss, and full protection of the service's own appropriation remains insufficient. Conversely, if every maximum floor reaches the discretionary reference, a rule-setter may expand at least one donor-backed service while preserving all the others. The model characterizes the complete set of such allocations and the minimum protection rates that implement them.

This characterization also closes the associated welfare problem. Within the preservation set, strict concavity yields a unique welfare-maximizing reform. A logarithmic specification supplies an equilibrium foundation for the fixed-gift analysis: sufficiently high donor valuations make full use of authorized funding capacities strictly optimal for every protection vector, while the authority strictly prefers accepting every gift. In that region, contribution capacities govern protectability and the order in which constraints bind, whereas heterogeneous donor valuations govern the welfare incidence of the induced reallocations.

A common protection rate provides a useful benchmark. Floors bind in descending order of gifts relative to pre-gift provision, so services with slack floors may lose provision while other floors tighten. Every donor-backed service is weakly nondecreasing along the entire common-rate path only when gift-to-baseline ratios coincide. A general incidence formula then shows how heterogeneous marginal valuations determine the welfare effect of the resulting fiscal transmission.

The paper contributes an applied public-finance characterization of baseline-referenced protection in a shared budget. Fiscal fungibility, preference heterogeneity, voluntary provision, and regulatory burden shifting have established foundations. Our object of analysis is the protection slack created by the difference between a service's contractual ceiling and its pooled post-gift reference. Its comparative statics identify a cross-donor fiscal externality, and the feasibility theorem establishes exactly when baseline protection is strong enough to deliver portfolio preservation. Taxes, pre-gift appropriations, enforceability, and the budget authority's objective remain predetermined. Public matching above the original appropriation expands the instrument set and leads to a different policy problem.

The remainder of the paper relates the argument to the literature, develops the shared-budget environment, derives the protectability and welfare results, and then establishes the voluntary-giving foundation and the incidence of common protection.

\section{Related Literature}

The paper relates first to work on heterogeneous preferences and the joint public and private provision of public goods. \citet{Epple2003} endogenize public financing and the permission of private contributions under collective choice. \citet{Bellani2015} show that heterogeneity in preferences over several public goods affects both the level and composition of in-kind redistribution, while \citet{Bellani2020} document a negative association between preference heterogeneity and several categories of public spending across countries. \citet{Ghosh2007} study heterogeneous contributors to several public goods when contributions may be bundled into a common pool before being allocated across the goods. Their mechanism concerns the contribution and welfare effects of bundling. Here gifts are purpose-specific, while the budget authority reallocates its own spending around them; the object is the fiscal benefit that baseline-referenced commitments are able to preserve. Our analysis therefore takes the budget authority's political objective as given and studies how heterogeneous donor purposes interact through a shared fiscal constraint without introducing a new aggregation rule.

The protection instrument has direct counterparts in maintenance-of-effort requirements and in rules requiring external funds to supplement rather than supplant domestic spending. Such rules seek to preserve the recipient's own spending when external funds arrive. \citet{GAO1980} emphasizes that maintenance-of-effort requirements may still leave room for fiscal substitution and may shift pressure toward other budget priorities. Current U.S. law provides a direct supplement-not-supplant formulation for specified federal education funds \citep{USCode7120}. In the philanthropic setting, the UK government's recent place-based strategy states that philanthropy should complement public money rather than substitute for state funding \citep{DCMS2026}. The same document uses ``additionality'' in a separate sense concerning the government's respect for donor choice, so our terminology should be read as a model definition rather than as a claim about uniform policy usage. The present paper formalizes a multi-service implication of the non-displacement principle: preservation of a pre-gift appropriation may still leave a service exposed to the loss of fiscal benefits created by other gifts.

The closest analytical fiscal antecedent is the literature on earmarked aid and expenditure displacement. \citet{Feyzioglu1998} analyze aid fungibility across public spending categories in a model with several public goods and specified fungibility parameters. Their analysis establishes why a targeted transfer must be evaluated within the entire public budget. We focus on enforceable expenditure floors linked to pre-gift appropriations and characterize which post-gift allocations remain attainable under those floors. The preservation test therefore compares the maximum contractual floor with the allocation after all contributions have entered the budget.

Gifts to public authorities provide another direct comparison. \citet{Slemrod2023} study voluntary gifts to government, tax evasion, and earmarking. Their two-public-good extension uses warm-glow benefits and fixed allocation proportions for taxes and unconditional gifts. In our model, donors value actual service provision and anticipate the authority's fiscal reallocation. This difference creates the channel through which heterogeneous donor purposes affect welfare after pooled fiscal responses.

Commitment and voluntary contributions also have established treatments. \citet{Boadway2007} analyze conditional and matching commitments in public-good contribution games, while \citet{Singhal2008} develops a dynamic explanation for incomplete fungibility when fundraising groups interact with political decision makers. Recent contributions in this journal study related interactions between private giving and public finance. \citet{HagenRattso2024} show how government grants alter fundraising incentives across competing charities, and \citet{MorenoTorregrosa2025} analyze the coexistence of voluntary contributions and tax finance for a public good. \citet{Zanoutene2026} studies the optimal mix of direct government grants and tax incentives for charitable giving across causes, showing that the ability to differentiate policy by cause affects the funding mix. Our model instead fixes the public budget envelope and baseline allocation before the protection problem and asks which post-gift service levels remain protectable when designated gifts share that budget. \citet{Chan2020} further distinguish contribution volume from the cost-effectiveness of provision across several public goods. Equal unit costs here remove productivity differences and isolate fiscal redistribution across purposes.

Finally, the relation to regulation helps delimit the mechanism. \citet{Che2000} study contracts with budget-constrained buyers, and the one-service version of our earlier protection problem admits an equivalent price representation. \citet{Amador2022} analyze transfer-free regulation with producer participation and identify conditions favoring price caps. In the shared-budget environment, services are coupled through a common resource constraint, so independent one-service pricing problems do not recover the allocation. Generic cross-market responses under regulation are themselves familiar, including the waterbed mechanism studied by \citet{Genakos2011}. The contribution here is the exact protectability condition generated by baseline-referenced fiscal commitments and its equilibrium implementation with heterogeneous donors.

\section{Environment and Reference Allocations}
\subsection{Public Services and the Protection Instrument}
There are $n\geq2$ donor-backed services. Public spending on service $i$ is $g_i$, its private contribution is $d_i$, and total provision is $q_i=g_i+d_i$. The public budget is $B>0$, and $X$ denotes expenditure on other services. One unit of provision costs one unit of a numeraire. The authority's objective is
\begin{equation}
 M(q,X)=\sum_{i=1}^n F_i(q_i)+H(X),
 \qquad X+\sum_i g_i=B.
 \label{eq:public}
\end{equation}
Each $F_i$ and $H$ is increasing and twice continuously differentiable, with a strictly negative second derivative on the relevant positive domain. The objective $M$ is the political objective of the constituency or authority controlling the public budget. It excludes donor utility, which enters the social-welfare criterion separately in Section~\ref{sec:welfare}. This separation avoids double counting and leaves the public budget objective exogenous rather than deriving it from a general voting model.

Let $(q^0,X^0)$ be the unique interior no-gift optimum and let $\lambda^0$ be its resource multiplier. Thus
\begin{equation}
 F_i'(q_i^0)=H'(X^0)=\lambda^0,
 \qquad \sum_iq_i^0+X^0=B.
 \label{eq:baseline}
\end{equation}
These appropriations are fixed before agreements are offered. Donor $i$ has capacity $w_i$ satisfying $0<w_i<q_i^0$ and chooses $d_i\in[0,w_i]$. The bound on capacity keeps own public spending positive and avoids financing corners unrelated to the protection question. Donor $i$ values its own service and incurs the contribution cost; its benefit function is specified below for the welfare and voluntary-giving results.

A rule-setter specifies componentwise minimum rates $s_i\in[0,1]$. The rule-setter may represent a legislature, an oversight body, or an enforceable funding agreement that constrains subsequent budget decisions. For an admitted agreement, the required public-spending floor is
\begin{equation}
 g_i\geq q_i^0-(1-s_i)d_i,
 \qquad\text{equivalently}\qquad
 q_i\geq q_i^0+s_id_i.
 \label{eq:instrument}
\end{equation}
At $s_i=1$, the original public appropriation is preserved. The restriction $s_i\leq1$ excludes rules that require public spending above that appropriation and thereby defines the instrument class. Public matching may be available in other institutional settings. Moreover, designating the recipient service leaves subsequent public expenditure unconstrained, whereas the protection rule imposes an explicit expenditure floor.

The rule-setter fixes the minimum requirements before the authority decides which agreements to admit. The authority may reject any agreement and may promise rates above the minima. Accepted donors then choose their contributions simultaneously, after which the authority reallocates its budget subject to the commitments. Rejected donors contribute zero. Agreement costs are zero, preferences and capacities are observable, commitments are enforceable, and independent private provision of an equivalent expansion is unavailable. The model takes the rule-setter's authority to impose these standards as an institutional feature; it does not derive that authority from the budget authority's own objective.

\subsection{A Fixed-Gift Comparison and Its Equilibrium Foundation}
Sections~\ref{sec:limit}--\ref{sec:welfare} initially condition on every gift reaching its capacity. Define total gifts $D=\sum_iw_i$ and total resources $R=B+D$. Let $(\qb,\Xb)$ denote the interior allocation at these resources when the authority retains full discretion. Its multiplier is $\lb$, so
\begin{equation}
 F_i'(\qb_i)=H'(\Xb)=\lb,
 \qquad\sum_i\qb_i+\Xb=R.
 \label{eq:reference}
\end{equation}
Strict concavity and the increased resources imply $\lb<\lambda^0$, $\qb_i>q_i^0$, and $\Xb>X^0$. Define
\begin{equation}
 L_i(s_i)=q_i^0+s_iw_i,
 \qquad u_i=q_i^0+w_i,
 \qquad h_i=u_i-\qb_i.
 \label{eq:room}
\end{equation}
Here $L_i$ is the imposed floor, $u_i$ the largest permissible floor, and $h_i$ its distance from discretionary post-gift provision. The quantity $u_i$ limits the contractual guarantee. The authority may still provide more voluntarily.

To keep the policy references distinct, we use \emph{baseline additionality} for preservation of the pre-gift public appropriation in a donor-backed service. A rate $s_i=1$ guarantees this property because it requires $g_i\geq q_i^0$. We use \emph{portfolio preservation} for the stronger requirement $q_i\geq\qb_i$, where $\qb_i$ is the service's provision after all gifts have entered the shared budget under full discretion. Baseline additionality may therefore hold while portfolio preservation fails. Portfolio preservation serves as a diagnostic benchmark for the reach of the instrument. It does not assign a distributive claim to $\qb_i$; in the welfare analysis it enters only as the maintained policy constraint whose implementable set is being characterized.

The relevant comparison uses the same $B$ and the same vector $w$ before and after a protection reform. A rule is \emph{effective} if it changes actual provision from $(\qb,\Xb)$. It \emph{preserves all donor-backed services} if $q_i\geq\qb_i$ for every $i$, with at least one strict inequality. This definition leaves $X$ free to decline. Section~\ref{sec:giving} supplies a voluntary-giving region in which the fixed-gift comparison is an equilibrium comparison under every rate vector. Outside that region, the allocation theorems remain conditional on unchanged gifts.

\begin{lemma}[Allocation and the common multiplier]\label{lem:allocation}
Let $f_i=(F_i')^{-1}$ and $x=(H')^{-1}$ on the relevant multiplier interval. At a fixed gift vector, the authority's unique final allocation satisfies
\begin{equation}
 q_i=\max\{L_i,f_i(\lambda)\},\qquad X=x(\lambda),
 \qquad \sum_iq_i+X=R,
 \label{eq:allocation}
\end{equation}
with $\lb\leq\lambda\leq\lambda^0$. A rule is effective exactly when some $L_i>\qb_i$. Every effective rule has $\lambda>\lb$ and $X<\Xb$.
\end{lemma}

The common multiplier links all services through the public budget. When one floor binds more tightly, the marginal value of public resources rises and the authority reduces provision in services that remain adjustable. Consequently, preservation depends on the level of the guaranteed floor relative to $\qb_i$, while the percentage rate alone is insufficient. All proofs are in the mathematical appendix.

\subsection{Cross-donor fiscal externalities in protectability}
The discretionary reference itself changes with the gift portfolio. Evaluate curvature at $(\qb,\Xb)$ and define
\begin{equation}
 \bar\tau_i=-\frac{1}{F_i''(\qb_i)},\qquad
 \bar\tau_X=-\frac{1}{H''(\Xb)},\qquad
 \bar T=\bar\tau_X+\sum_k\bar\tau_k.
 \label{eq:referencecurvature}
\end{equation}
These terms describe how the authority spreads a marginal increase in the pooled resources across its unrestricted portfolio.

\begin{proposition}[Cross-donor protectability]\label{prop:protectability}
Hold $B$, public preferences, and the pre-gift allocation fixed, and vary one realized gift $w_j$ locally. At the discretionary post-gift allocation,
\begin{equation}
 \frac{\partial\qb_i}{\partial w_j}=\frac{\bar\tau_i}{\bar T}>0,\qquad
 \frac{\partial\Xb}{\partial w_j}=\frac{\bar\tau_X}{\bar T}>0.
 \label{eq:giftpass}
\end{equation}
Consequently, the protection slack $h_i=q_i^0+w_i-\qb_i$ satisfies
\begin{equation}
 \frac{\partial h_i}{\partial w_i}=1-\frac{\bar\tau_i}{\bar T}>0,\qquad
 \frac{\partial h_i}{\partial w_j}=-\frac{\bar\tau_i}{\bar T}<0\quad(j\ne i).
 \label{eq:protectabilityderivatives}
\end{equation}
A larger own gift therefore makes service $i$ easier to preserve within the baseline-referenced instrument class, whereas a larger gift to another purpose makes it harder to preserve.
\end{proposition}

Every gift raises the discretionary provision of every service because it enlarges the pooled resource envelope. The recipient service simultaneously gains a one-for-one increase in its own maximum contractual floor. Other services receive the fiscal benefit without any corresponding increase in their contractual ceilings. Equation~\eqref{eq:protectabilityderivatives} therefore identifies a cross-donor fiscal externality in protectability. The effect depends on public-budget curvature, while donor valuations enter later through giving incentives and welfare incidence.

\section{Baseline Additionality and Portfolio Preservation}\label{sec:limit}
\begin{theorem}[Unavoidable losses and the complete preservation set]\label{thm:main}
Under the assumptions of Section~3, conditional on gifts $d_i=w_i$:

\noindent\textup{(i)} If $h_j<0$ for a service $j$, every effective reform generated by any $s\in[0,1]^n$ satisfies $q_j<\qb_j$. This includes reforms with $s_j=1$.

\noindent\textup{(ii)} An effective reform preserving all donor-backed services exists if and only if
\begin{equation}
 q_i^0+w_i\geq\qb_i\qquad\text{for every }i.
 \label{eq:criterion}
\end{equation}
When this condition holds, the complete set of such allocations is
\begin{equation}
 q_i=\qb_i+t_i,\qquad 0\leq t_i\leq h_i,\qquad
 \sum_it_i>0,\qquad X=\Xb-\sum_it_i.
 \label{eq:box}
\end{equation}
Each is implemented by the minimum rates
\begin{equation}
 s_i=\frac{\qb_i-q_i^0+t_i}{w_i}\in[0,1].
 \label{eq:rates}
\end{equation}
Every donor-backed service floor binds at the implemented allocation and $X\geq X^0>0$.
\end{theorem}

The negative statement applies service by service. If several maximum floors lie below their discretionary references, each of those services loses under every effective rate vector. The rates may differ across services, and the conclusion does not depend on which floor first excludes the discretionary allocation. For a vulnerable service, its largest permissible floor and its desired unconstrained quantity at the higher multiplier both lie below its previous provision. Full protection of its own appropriation therefore remains insufficient.

The positive statement uses the budget's remaining flexibility differently. Set every service floor at its discretionary post-gift provision and add feasible increments. All donor-backed services are then insulated, and $X$ finances the expansion. The identity
\begin{equation}
 \sum_i h_i=\Xb-X^0>0
 \label{eq:sumroom}
\end{equation}
ensures that a positive increment exists whenever every $h_i$ is nonnegative. Hence the weak inequalities in~\eqref{eq:criterion} are sufficient; an additional strict inequality assumption is unnecessary.

Condition~\eqref{eq:criterion} is a limitation of the permitted commitments. A service with $h_j<0$ would need a floor of at least $\qb_j$ to survive any effective reform without a reduction. In the original parametrization the required rate would be
\begin{equation}
 s_j^{\mathrm{preserve}}=\frac{\qb_j-q_j^0}{w_j}>1.
 \label{eq:aboveone}
\end{equation}
At that floor, required public spending exceeds the pre-gift appropriation by at least $\qb_j-q_j^0-w_j=-h_j$. The expression identifies the additional public contribution needed for that service; it makes no assumption about how the authority would finance it. A richer matching system therefore changes both the feasible set and the relevant incentive problem.

The theorem also clarifies the single-service benchmark. With one donor-backed service and a positive residual category, identity~\eqref{eq:sumroom} gives $h_1>0$, so every effective floor raises that service and lowers $X$. The impossibility associated with a small own gift therefore arises from cross-service fiscal interactions. Even a service that loses under the multi-service rule still receives at least $q_j^0$; the relevant loss is measured relative to discretionary post-gift provision.

\section{Welfare within the Preservation Set}\label{sec:welfare}
Let $V_i$ be differentiable and concave donor benefit functions satisfying $V_i'(q)>0$ on the relevant domain. Donors are outside the political objective $M$ in~\eqref{eq:public}; their utilities enter the welfare criterion here. Using comparable utility units and equal marginal values of the numeraire, total welfare is
\begin{equation}
 W(q,X)=\sum_i[F_i(q_i)+V_i(q_i)]+H(X)-D.
 \label{eq:welfare}
\end{equation}
The constant public financing cost has been omitted. At unchanged gifts, a service-preserving reform weakly improves every donor's payoff. The authority, by contrast, loses from any effective reform because $(\qb,\Xb)$ uniquely maximizes its objective. Section~\ref{sec:giving} later establishes strict agreement acceptance relative to rejection, which is a separate comparison.

If condition~\eqref{eq:criterion} holds, choose a service $k$ with $h_k>0$ and increase only that service by a small amount $\varepsilon$. At the discretionary allocation,
\begin{equation}
 \left.\frac{dW}{d\varepsilon}\right|_{\varepsilon=0}
 =F_k'(\qb_k)-H'(\Xb)+V_k'(\qb_k)
 =V_k'(\qb_k)>0.
 \label{eq:localwelfare}
\end{equation}
The authority's first-order loss is zero, while the donor gains to first order. Hence sufficiently small service-preserving reforms increase total welfare. This is a local statement; large expansions may impose substantial costs through reduced $X$.

\subsection{Choosing the Welfare-Best Implementable Reform}
The exact set in Theorem~\ref{thm:main} also makes the policy choice explicit. Define the closed set $\mathcal T=\prod_i[0,h_i]$ and the objective
\begin{equation}
 \Phi(t)=\sum_i[F_i(\qb_i+t_i)+V_i(\qb_i+t_i)]
       +H\!\left(\Xb-\sum_i t_i\right)-D.
 \label{eq:policyobjective}
\end{equation}
Its domain has $X\geq X^0$. The authority is still the final allocator; choosing $t$ here means implementing it with the minimum rates~\eqref{eq:rates}.

\begin{corollary}[Best policy preserving all donor-backed services]\label{cor:best}
If~\eqref{eq:criterion} holds, $\Phi$ has a unique maximizer $t^*$ on $\mathcal T$. It is nonzero and therefore implementable as an effective reform. Let $X^*=\Xb-\sum_it_i^*$ and define the net marginal social value
\begin{equation}
 \Delta_i=F_i'(\qb_i+t_i^*)+V_i'(\qb_i+t_i^*)-H'(X^*).
 \label{eq:marginal}
\end{equation}
For every coordinate with $h_i>0$, the necessary and sufficient optimality conditions are
\begin{equation}
 \begin{cases}
 \Delta_i\leq0,&t_i^*=0,\\
 \Delta_i=0,&0<t_i^*<h_i,\\
 \Delta_i\geq0,&t_i^*=h_i.
 \end{cases}
 \label{eq:policykkt}
\end{equation}
Coordinates with $h_i=0$ are fixed. The rule maximizes welfare among implementable allocations preserving all donor-backed services, not among all possible fiscal instruments or all protection reforms.
\end{corollary}

Strict concavity supplies uniqueness, and equation~\eqref{eq:localwelfare} rules out the zero vector. These standard optimality conditions close the policy problem on the exact feasible set. If condition~\eqref{eq:criterion} fails, the preservation set contains no nonzero reform. A rule-setter that permits losses in individual services faces a different welfare problem, and Theorem~\ref{thm:main} leaves the sign of total welfare for such reforms open.

\section{Voluntary Giving and Government Acceptance}\label{sec:giving}
The preceding geometry conditions on the gift vector. This section establishes one environment in which that vector is chosen voluntarily and remains unchanged for every policy used above. Specify
\begin{equation}
 M(q,X)=\sum_i a_i\ln q_i+b\ln X,
 \qquad U_i(q_i,d_i)=\theta_i\ln q_i-d_i,
 \label{eq:logutility}
\end{equation}
with $a_i,b,\theta_i>0$. Put $A=b+\sum_i a_i$. Baselines and the discretionary post-gift allocation are
\begin{equation}
 q_i^0=\frac{a_iB}{A},\quad X^0=\frac{bB}{A},\qquad
 \qb_i=\frac{a_i(B+D)}{A},\quad\Xb=\frac{b(B+D)}{A}.
 \label{eq:logreferences}
\end{equation}
Maintain $0<w_i<q_i^0$.

\begin{lemma}[Full giving and strict admission under every rate vector]\label{lem:giving}
If
\begin{equation}
 \theta_i>B+D\qquad\text{for every donor }i,
 \label{eq:highvaluation}
\end{equation}
contributing $w_i$ is strictly dominant for every admitted donor under every $s\in[0,1]^n$, every admission subset, and every feasible vector of other contributions. The authority strictly prefers adding each missing gift. It therefore admits all agreements. Choosing actual protection rates at the componentwise required minima is optimal; higher slack rates may be payoff-equivalent. Final allocations and the contribution vector are unique.
\end{lemma}

The sufficient bound is deliberately strong, yet it describes an open parameter region and verifies giving even away from equilibrium. A donor whose floor is slack expands the resources available to the freely adjusted categories, whereas a donor whose floor binds raises that floor. In both cases, the logarithmic marginal benefit from its own contribution is bounded below by $\theta_i/(B+D)$. Donor utility therefore rises throughout the authorized funding interval.

Admission is strict. Starting from any admitted subset, the authority may add a missing donor's entire capacity to that donor's service while leaving all other quantities unchanged. The resulting allocation is feasible under every rate at most one and strictly increases the authority's benefit, while the other admitted donors continue to give their capacities. Because higher promised rates attract no additional finance and only shrink the feasible allocation set, the authority optimally implements the required minima.

The authority accepts because receiving the gift under the rule is preferable to rejecting it under the same imposed rule. It may still prefer unrestricted discretion to the imposed rule itself. Likewise, the sufficient bound does not imply that donors generally exhaust an exogenous budget. A pre-authorized grant envelope provides one interpretation of capacity, while its optimal determination lies outside the model. Warm glow, private information, and contributions outside the high-valuation region may alter the response.

\begin{corollary}[Donation shares and preservable provision]\label{cor:shares}
Under Lemma~\ref{lem:giving}, a nontrivial reform preserving every donor-backed service exists exactly when
\begin{equation}
 w_i\geq\frac{a_iD}{A}\quad\text{for all }i,
 \qquad\text{equivalently}\qquad
 \frac{w_i}{D}\geq\frac{a_i}{A}\quad\text{for all }i.
 \label{eq:shares}
\end{equation}
Any service violating this inequality loses provision under every effective rate vector, even with its own full protection. All gifts remain at capacity and all agreements remain strictly accepted.
\end{corollary}

The authority's share for a service is measured relative to all categories, including $X$. Hence all inequalities may hold simultaneously. A small gift relative to that share leaves the service dependent on the fiscal benefits of others' contributions. Its own maximum commitment may then be insufficient to preserve those benefits. Proposition~\ref{prop:protectability} simplifies in this logarithmic environment to
\begin{equation}
 \frac{\partial h_i}{\partial w_i}=1-\frac{a_i}{A},\qquad
 \frac{\partial h_i}{\partial w_j}=-\frac{a_i}{A}\quad(j\ne i).
 \label{eq:logprotectability}
\end{equation}
Thus the public expenditure weight determines how strongly another donor's contribution erodes service $i$'s protection slack.

For logarithmic donors, the welfare calculation in Corollary~\ref{cor:best} has an especially convenient representation. If~\eqref{eq:shares} holds, its unique solution satisfies
\begin{equation}
 q_i^*=\min\!\left\{u_i,\max\!\left\{\qb_i,\frac{a_i+\theta_i}{\mu}\right\}\right\},
 \quad X^*=\frac b\mu,\quad \sum_iq_i^*+X^*=B+D,
 \label{eq:welfarewaterfill}
\end{equation}
where $\mu>0$ is the social resource multiplier. Substituting $t_i^*=q_i^*-\qb_i$ in~\eqref{eq:rates} gives the implementing rates. Since $X^*>0$, the scalar resource equation is strictly decreasing in $\mu$ and therefore has a unique solution.

\subsection{An Exact Two-Service Illustration}
Consider the logarithmic environment with $B=30$, $a_1=a_2=b=10$, $w_1=5$, $w_2=1$, $\theta_1=40$, and $\theta_2=100$. Both valuations exceed $B+D=36$, so Lemma~\ref{lem:giving} applies. The no-gift allocation is $(10,10,10)$, discretionary post-gift provision is $(12,12,12)$, and the maximum service floors are $(15,11)$. Thus $h=(3,-1)$: the second service fails to preserve its discretionary fiscal benefit under every effective protection reform. For example, minima $(s_1,s_2)=(0.6,1)$ yield $(q_1,q_2,X)=(13,11.5,11.5)$. Even full own protection guarantees only eleven units for the second service, below its discretionary reference of twelve.

The positive side of Theorem~\ref{thm:main} is equally transparent. Raising only $w_2$ to four gives discretionary provision $(13,13,13)$ and maximum floors $(15,14)$, hence $h=(2,1)$. The preservation set is then nonempty. These exact calculations clarify the two reference levels and the instrument ceiling, while the theorem itself is independent of the selected values.

\section{Common Protection and Cross-Service Incidence}
Differentiated rates expand the rule-setter's choice set relative to a common rate. The uniform benchmark remains informative because it shows how contribution capacities determine which floors bind first, whereas donor valuations affect the welfare consequences of the resulting reallocation.

\subsection{Binding Order and the Proportionality Exception}
In the logarithmic full-giving region let $s_i=s$ and define $r_i=w_i/q_i^0$. If $I$ denotes the services with strictly binding floors on an interval, put $D_I=\sum_{i\in I}w_i$ and $A_F=b+\sum_{j\notin I}a_j$. The final allocation has
\begin{equation}
 q_i(s)=\max\!\left\{q_i^0+sw_i,\frac{a_i}{\lambda(s)}\right\},
 \quad X(s)=\frac b{\lambda(s)},\quad \sum_iq_i(s)+X(s)=B+D.
 \label{eq:commonallocation}
\end{equation}

\begin{proposition}[Common-rate incidence]\label{prop:common}
Floors bind in descending order of $r_i$, simultaneously for tied ratios, and remain binding thereafter. The first binding rate is
\begin{equation}
 s_0=\frac{D}{B\max_i r_i}<1.
 \label{eq:firstbinding}
\end{equation}
Within an interval with fixed strict binding set,
\begin{equation}
 \frac{dq_i}{ds}=w_i\ (i\in I),\qquad
 \frac{dq_j}{ds}=-\frac{a_jD_I}{A_F}\ (j\notin I),\qquad
 \frac{dX}{ds}=-\frac{bD_I}{A_F}.
 \label{eq:commonderivatives}
\end{equation}
Every donor-backed service is weakly nondecreasing over the whole path $s\in[0,1]$ if and only if all $r_i$ are equal. Otherwise at least one service strictly declines on a nonempty interval.
\end{proposition}

The common-rate condition is stronger than~\eqref{eq:shares}. Differentiation may preserve all services even when a common rate causes losses, provided the maximum floors are sufficiently high. Conversely, services below the threshold in~\eqref{eq:shares} lose under every effective vector, so abandoning uniformity does not protect them. The proportionality condition describes componentwise provision throughout the common path; it says nothing by itself about welfare monotonicity.

\subsection{A General Welfare-Incidence Formula}
For the local incidence calculation, return to general public preferences and keep contributions fixed. Let $I$ be a fixed strict binding set under a common rate. Define positive curvature measures
\begin{equation}
 \tau_j=-\frac1{F_j''(q_j)}\ (j\notin I),\qquad
 \tau_X=-\frac1{H''(X)},\qquad T=\tau_X+\sum_{j\notin I}\tau_j.
 \label{eq:curvatures}
\end{equation}

\begin{proposition}[Fiscal transmission and social incidence]\label{prop:incidence}
On such an interval,
\begin{equation}
 \lambda'=\frac{D_I}{T},\qquad q_j'=-\frac{\tau_jD_I}{T}\ (j\notin I),\qquad
 X'=-\frac{\tau_XD_I}{T}.
 \label{eq:generalderivatives}
\end{equation}
With welfare~\eqref{eq:welfare},
\begin{equation}
 W'=\sum_{i\in I}w_i\bigl[F_i'(q_i)+V_i'(q_i)-\lambda\bigr]
       -D_I\sum_{j\notin I}\frac{\tau_j}{T}V_j'(q_j).
 \label{eq:welfareincidence}
\end{equation}
At active-set changes the relevant derivatives are one-sided.
\end{proposition}

The final term captures the donor benefit lost in services that finance the protected expansion. A one-service model with an unvalued residual category omits this channel. Equation~\eqref{eq:welfareincidence} therefore provides a local test for the welfare effect and leaves its sign to the underlying preferences. For logarithmic preferences and a uniquely first-binding service $i$, write $\lambda^u=A/(B+D)$. At the first binding point,
\begin{equation}
 W'(s_0+) = w_i\lambda^u\left[\frac{\theta_i}{a_i}
             -\frac{\sum_{j\ne i}\theta_j}{b+\sum_{j\ne i}a_j}\right].
 \label{eq:harmcondition}
\end{equation}
A negative bracket implies a local welfare loss despite unchanged giving and strict admission. At that initial point the authority's first-order loss is zero; the sign is determined by the donor values attached to the reallocations.

\section{Interpretation, Scope, and the Single-Service Benchmark}
The model distinguishes preservation of an appropriation from preservation of the fiscal benefits generated by pooled gifts. The first promise may be too weak to deliver the second because its ceiling is tied to the service's own contribution, whereas the reference allocation depends on every contribution. This distinction drives the impossibility result and explains when differentiated protection is effective.

The conclusion depends on a fixed public budget and stable contributions. The discretionary reference is a counterfactual used to identify which services finance a reform; it carries no claim of distributive entitlement. Different welfare weights, benefits to populations outside the model, or alternative policy objectives may justify reducing a service. The preservation constraint therefore represents one policy objective among several.

The rule-setter is institutionally distinct from the budget authority. A legislature, oversight body, or funding agreement may impose the minimum requirement even though the authority would prefer greater discretion. Once the minima are in place, the authority optimally accepts gifts in the equilibrium region studied above. Enforcement costs, strategic changes in appropriations, and subsidies may alter both the adoption of such standards and the feasible set.

The former one-service pricing approach remains a useful boundary case. With linear residual utility and a baseline satisfying $F'(q^0)=1$, one obtains $q-q^0=sd$, so $p=1/s$ gives the implicit price of additional provision and a minimum rate acts as a price cap. Under a shared budget, a slack floor may satisfy $q_i-q_i^0>s_id_i$, and its quantity then responds to other gifts and other floors. Independent scalar price problems therefore no longer recover the allocation.

This failure does not eliminate standard representations. At fixed gifts define
\begin{equation}
 \mathcal C(q)=M(\qb,\Xb)-\left[\sum_i F_i(q_i)+H\!\left(R-\sum_iq_i\right)\right].
 \label{eq:coupledcost}
\end{equation}
The authority minimizes this convex cost subject to the floors. Its Hessian is
\begin{equation}
 \nabla^2\mathcal C(q)=\operatorname{diag}(-F_i''(q_i))-H''(X)\mathbf1\mathbf1^{\mathsf T},
 \label{eq:costhessian}
\end{equation}
which is positive definite with positive cross-derivatives. This representation links the model to standard coupled allocation problems. The economic contribution lies in the protectability externality, the exact feasibility boundary for baseline-referenced commitments, and its voluntary-giving implementation.

Finally, the endogenous-giving foundation is intentionally limited. If donors leave capacity unused, a protected expansion may change the contributions of other donors as well as the allocation of public spending. The main theorems isolate the fiscal mechanism by establishing capacity exhaustion for every policy in the stated region. Endogenous contributions outside that region, private information, and nonseparable public preferences require separate analysis.

\section{Conclusion}

Heterogeneous philanthropic preferences give private funding a natural role in directing resources toward particular public services, while the budget authority continues to allocate its own resources across the entire portfolio. This interaction makes additionality depend on both the gift and the fiscal response. A commitment to preserve pre-gift public funding therefore protects a different object from the service level generated after all gifts are pooled.

The model separates these objects as baseline additionality and portfolio preservation. A larger own gift increases a service's protection slack because its contractual ceiling rises faster than its discretionary reference. Gifts to other purposes have the opposite effect: they raise the discretionary reference without raising that service's ceiling. This cross-donor fiscal externality explains why a rule that successfully prevents direct displacement may still expose a donor-backed service to losses relative to pooled discretionary budgeting.

The feasibility boundary is exact. If a service's maximum permissible floor lies below its discretionary post-gift provision, every effective reform in the permitted class reduces that service relative to the portfolio reference. When every maximum floor reaches the reference, differentiated rates implement the complete preservation set, and strict concavity selects a unique welfare-best element. In the logarithmic equilibrium, donation shares determine protectability while heterogeneous donor valuations determine the social incidence of the reallocations.

The policy implication concerns the benchmark embedded in an additionality rule. Maintenance-of-effort and supplement-not-supplant arrangements protect a baseline level of public commitment. In a shared budget, that baseline may fall short of the fiscal benefit a service receives from the full portfolio of gifts. Preserving that broader reference requires sufficiently large own contributions, revised reference levels, public matching, or another instrument that expands the contractual ceiling. The appropriate design therefore depends on which notion of additionality the institution intends to secure.

\section*{Declarations}
\noindent\textbf{Data availability.} The article uses no empirical data. The numerical illustration is generated from the theoretical model.

\noindent\textbf{Funding.} No specific external funding was received for this research. The work was undertaken within the author's affiliated organizations.

\noindent\textbf{Conflict of interest.} The author declares no conflict of interest.

\noindent\textbf{Use of artificial intelligence.} During the preparation of this work, the author used ChatGPT (OpenAI) for support in reviewing theoretical arguments, scrutinizing derivations, and editing the manuscript.

\appendix

\section{Proofs of the Allocation and Policy Results}
\subsection{The Multiplier and Effective Reforms}
\begin{proof}[Proof of Lemma~\ref{lem:allocation}]
The unrestricted first-order conditions equate all marginal benefits. Since every inverse marginal-benefit function decreases strictly, an increase from $B$ to $R=B+D$ lowers the resource multiplier and increases every category. This yields $\lb<\lambda^0$, $\qb_i>q_i^0$, and $\Xb>X^0$.

The allocation with $q_i=L_i$ and $X=R-\sum_iL_i\geq X^0$ is feasible. At multiplier $\lb$, required resources are
\[
 \sum_i\max\{L_i,\qb_i\}+\Xb
 = R+\sum_i(L_i-\qb_i)_+\geq R.
\]
At $\lambda^0$, required resources are $\sum_iL_i+X^0\leq R$, because $L_i\geq q_i^0$ and $L_i\leq q_i^0+w_i$. The resource-demand function is continuous and strictly decreasing on this interval, with strictness supplied by the residual category. Hence it has a unique root. The lower-bound first-order conditions give~\eqref{eq:allocation}; strict concavity makes it the unique optimum. Public spending is positive since $q_i\geq q_i^0>w_i$.

If every $L_i\leq\qb_i$, the discretionary optimum remains feasible and optimal. If some $L_i>\qb_i$, required resources at $\lb$ exceed $R$, so the new multiplier strictly exceeds $\lb$ and $X<\Xb$. This also excludes the old allocation and proves the equivalence.
\end{proof}

\subsection{Cross-Donor Protectability}
\begin{proof}[Proof of Proposition~\ref{prop:protectability}]
At the unrestricted post-gift allocation, differentiation of $F_i'(\qb_i)=\lb$ and $H'(\Xb)=\lb$ with respect to $w_j$ gives
\[
 d\qb_i=-\bar\tau_i\,d\lb,\qquad d\Xb=-\bar\tau_X\,d\lb.
\]
A marginal increase in $w_j$ raises total resources by the same amount. Differentiating the resource constraint therefore yields $-\bar T\,d\lb=dw_j$. Equation~\eqref{eq:giftpass} follows. Since $h_i=q_i^0+w_i-\qb_i$ and the pre-gift allocation is fixed, differentiation gives~\eqref{eq:protectabilityderivatives}. The residual category and the remaining public services imply $\bar T>\bar\tau_i$, so the own-gift derivative is strictly positive.
\end{proof}

\subsection{The Exact Preservation Set}
\begin{proof}[Proof of Theorem~\ref{thm:main}]
For an effective rule Lemma~\ref{lem:allocation} gives $\lambda>\lb$. If $h_j<0$, then $L_j\leq u_j<\qb_j$, while $f_j(\lambda)<f_j(\lb)=\qb_j$. Both arguments of the maximum in~\eqref{eq:allocation} are below $\qb_j$. This proves part~(i) and necessity in part~(ii).

Suppose every $h_i\geq0$. Resource accounting gives
\[
 \sum_i h_i=\sum_i(q_i^0+w_i-\qb_i)
 =(B-X^0)+D-(R-\Xb)=\Xb-X^0>0.
\]
Choose any $t$ in~\eqref{eq:box}. Define $L_i=\qb_i+t_i$ using~\eqref{eq:rates}. Since $q_i^0<\qb_i\leq L_i\leq u_i$, every required rate lies in $[0,1]$. The residual allocation $X=\Xb-\sum_it_i$ belongs to $[X^0,\Xb)$ and satisfies
\[
 H'(X)>H'(\Xb)=F_i'(\qb_i)\geq F_i'(L_i)
\]
for each service. Thus all lower-bound optimality conditions hold with equality in the service floors and the common resource constraint. Strict concavity establishes that this is the authority's unique optimum, proving sufficiency and implementation of every stated allocation.

Conversely, an effective rule preserving all services must have a strict increase in some service; otherwise the budget also fixes $X=\Xb$, contradicting effectiveness. Lemma~\ref{lem:allocation} gives $\lambda>\lb$. Any service whose floor were slack would satisfy $q_i=f_i(\lambda)<\qb_i$, a contradiction. Thus every service binds at $q_i=L_i\leq u_i$, which yields $0\leq t_i\leq h_i$ and the resource identity in~\eqref{eq:box}. This proves that the set is complete.
\end{proof}

\subsection{The Welfare-Maximizing Preservation Policy}
\begin{proof}[Proof of Corollary~\ref{cor:best}]
The feasible box $\mathcal T$ is compact and has a positive-width coordinate by~\eqref{eq:sumroom}. Its residual expenditure satisfies $X\geq X^0>0$, so $\Phi$ is continuous. It is strictly concave on the free coordinates: each $F_i+V_i$ is strictly concave and $H$ is concave in the affine residual. A unique maximum follows. At $t=0$, the derivative in any coordinate with $h_i>0$ equals $V_i'(\qb_i)>0$ by~\eqref{eq:reference}. Consequently the maximizer is nonzero. Its partial derivatives are exactly~\eqref{eq:marginal}. The stated endpoint and interior signs are the first-order conditions for a concave function on a box, and are sufficient as well as necessary. Theorem~\ref{thm:main} implements the solution.

For the logarithmic representation~\eqref{eq:welfarewaterfill}, maximize $\sum_i(a_i+\theta_i)\ln q_i+b\ln X-D$ subject to $\qb_i\leq q_i\leq u_i$ and resource balance. Its optimality conditions clip $(a_i+\theta_i)/\mu$ at those bounds and give $X=b/\mu$. The resource equation has a unique root because the term $b/\mu$ is strictly decreasing. This yields the same unique allocation as the box formulation.
\end{proof}

\section{Proofs for Endogenous Giving and Common Rules}
\subsection{Voluntary Giving for Arbitrary Differentiated Minima}
\begin{proof}[Proof of Lemma~\ref{lem:giving}]
Fix an admission subset and actual rate vector; let rejected donors' contributions be zero. Allow every other admitted contribution to vary in its feasible interval. Write $D_d=\sum_i d_i$. Logarithmic final allocation is
\[
 q_i=\max\{q_i^0+s_id_i,a_i/\lambda\},\qquad X=b/\lambda,
 \qquad \sum_iq_i+X=B+D_d.
\]
At $\lambda^0=A/B$, required resources are $B+\sum_i s_id_i\leq B+D_d$. Thus $\lambda\leq\lambda^0$, every service receives at least $q_i^0$, and own public funding remains positive.

Consider a smooth segment as admitted donor $j$ varies $d_j$. Let $I$ be the binding set and $A_F=b+\sum_{i\notin I}a_i$. If $j\notin I$, freely allocated resources are
\[
 R_F=B+D_d-\sum_{i\in I}(q_i^0+s_id_i).
\]
They increase by one per unit of $d_j$. Since $q_j=a_jR_F/A_F$,
\[
 \frac{\partial\ln q_j}{\partial d_j}=\frac1{R_F}\geq\frac1{B+D}.
\]
If $j\in I$ and $d_j>0$, define
\[
 K=\sum_{k\notin I}d_k+\sum_{k\in I\setminus\{j\}}(1-s_k)d_k\geq0,
 \qquad R_F=\frac{BA_F}{A}+K+(1-s_j)d_j.
\]
The binding condition $q_j^0+s_jd_j\geq a_jR_F/A_F$ implies
\[
 s_j(A_F+a_j)d_j\geq a_j(K+d_j),\qquad
 s_j\geq\frac{a_j}{A_F+a_j}.
\]
Therefore $q_j^0/s_j\leq B(A_F+a_j)/A\leq B$ and
\[
 \frac{\partial\ln q_j}{\partial d_j}
 =\frac{s_j}{q_j^0+s_jd_j}
 =\frac1{q_j^0/s_j+d_j}\geq\frac1{B+D}.
\]
There are finitely many active-set regions, with continuous allocations and one-sided limits at their boundaries. The same bound applies at zero by a limit. Donor utility is therefore strictly increasing throughout its interval when $\theta_j>B+D$. Giving $w_j$ is its unique best response independently of the other contributions.

For admission, compare any subset with that subset plus donor $j$. The old allocation has $q_j\geq q_j^0$. Add $w_j$ only to $q_j$ and leave all other quantities unchanged. This uses the extra resources, satisfies the new floor for every $s_j\leq1$, and strictly raises the authority's utility. Other admitted donors keep giving their capacities, so every missing agreement is strictly beneficial. The resulting strict subset comparisons establish full admission. At fixed contributions, choosing any vector above the componentwise minima only shrinks the feasible allocation set. The minimum vector is optimal. Higher slack floors may give the same unique allocation, so uniqueness of the promised rates is not claimed.
\end{proof}

\begin{proof}[Proof of Corollary~\ref{cor:shares}]
Equation~\eqref{eq:logreferences} gives $\qb_i-q_i^0=a_iD/A$. Substituting in~\eqref{eq:criterion} gives~\eqref{eq:shares}. Lemma~\ref{lem:giving} makes the fixed-gift theorem an equilibrium comparison under all rate vectors. The loss and strict-admission conclusions follow directly.
\end{proof}

\subsection{Binding Order and Common Protection}
\begin{proof}[Proof of Proposition~\ref{prop:common}]
Set $c=\lambda^0/\lambda\geq1$. The allocation is
\[
 q_i=q_i^0\max\{1+sr_i,c\},\qquad X=X^0c.
\]
For $s>0$ define $h=(c-1)/s$. Resource balance becomes
\begin{equation}
 \frac Ds=Bh+\sum_iq_i^0(r_i-h)_+.
 \label{eq:cutoff}
\end{equation}
The right side is strictly increasing in $h$, with piecewise derivative $B-\sum_{r_i>h}q_i^0\geq X^0>0$. Thus $h$ decreases as $s$ rises; the binding set $\{i:r_i>h\}$ expands in descending ratio order. Initially $c=1+D/B$, so the first binding point is~\eqref{eq:firstbinding}. Since $D\leq\max r_i\sum_iq_i^0<B\max r_i$, this rate is below one.

On a fixed binding interval the free categories share resources
\[
 R_F=B+D-\sum_{i\in I}(q_i^0+sw_i).
\]
Their logarithmic allocations are $q_j=a_jR_F/A_F$ and $X=bR_F/A_F$. Differentiation yields~\eqref{eq:commonderivatives}.

If not all ratios coincide, at least one floor is strictly slack when the highest-ratio floors first bind. Continuity gives a nonempty subsequent interval with positive $D_I$ and that service still free. Its derivative is strictly negative. If every ratio equals $r$, all floors first bind simultaneously at
\[
 s_0=\frac D{Br}=\frac{\sum_iq_i^0}{B}=1-\frac{X^0}{B}.
\]
Beforehand every service is unchanged; afterwards each has $q_i=q_i^0+sw_i$, which increases. At $s=1$ all positive-gift floors are strictly binding. Ties affect active-set labels at isolated boundaries, not the allocations or the conclusion.
\end{proof}

\subsection{General Curvature and Welfare Incidence}
\begin{proof}[Proof of Proposition~\ref{prop:incidence}]
For free services $F_j'(q_j)=\lambda$, and $H'(X)=\lambda$. Differentiation gives $q_j'=-\tau_j\lambda'$ and $X'=-\tau_X\lambda'$. Bound services have derivative $w_i$, so the fixed resource constraint implies $D_I-T\lambda'=0$. This proves~\eqref{eq:generalderivatives}. Differentiating~\eqref{eq:welfare}, using the free first-order conditions and the resource identity, yields
\[
 W'=\sum_{i\in I}(F_i'+V_i')w_i
       -\frac{D_I}{T}\left[\lambda\tau_X+
                 \sum_{j\notin I}(\lambda+V_j')\tau_j\right],
\]
which reduces to~\eqref{eq:welfareincidence}.

At the first binding point in the logarithmic model, all public marginal benefits equal $\lambda^u$. Free curvatures are $\tau_j=a_j/(\lambda^u)^2$ and $\tau_X=b/(\lambda^u)^2$. Donor marginal benefit is $V_j'=\theta_j\lambda^u/a_j$. Substitution for a single newly binding service gives~\eqref{eq:harmcondition}. Strict negativity persists under small primitive changes that preserve the first-binding order and the strict giving conditions.
\end{proof}

\section{The One-Service Price Representation}
With $M=X+F(q)$, the no-gift baseline satisfies $F'(q^0)=1$. A binding protection rule gives $q=q^0+sd$. For $s>0$, put $z=sd$ and $p=1/s$. The donor solves $\max_{z\geq0,pz\leq w}\{V(q^0+z)-pz\}$ and the authority's gain is $pz-C(z)$, where
\[
 C(z)=z-[F(q^0+z)-F(q^0)].
\]
The effective cost is convex and increasing above zero. A lower bound on $s$ is a price cap. This equivalence connects the benchmark to budget-constrained contracting and regulation \citep{Che2000,Amador2022}. The shared-budget analysis adds the multi-service protectability criterion to this scalar benchmark. Equation~\eqref{eq:coupledcost} provides a standard nonseparable representation of the coupled allocation problem.

\end{document}